\documentclass[a4wide,11pt,english]{article}
\pdfoutput=1
\usepackage[T1]{fontenc}
\usepackage[utf8]{inputenc}
\usepackage{geometry}
\usepackage{amssymb}
\usepackage{graphicx}
\usepackage{subcaption}
\usepackage{graphics}
\usepackage{amsmath}
\numberwithin{equation}{section}

\usepackage{chngcntr}
\counterwithin{figure}{section}
\usepackage{esint}
\usepackage{cite}
\usepackage{amsmath}
\usepackage{feynmp}
\usepackage{mathrsfs}
\usepackage{braket}
\usepackage{dsfont}
\usepackage{cancel}
\usepackage[utf8]{inputenc}
\usepackage{color}

\makeatletter
\providecommand{\tabularnewline}{\\}
\makeatother
\usepackage{babel}
\usepackage{hyperref}
\def\equationautorefname~#1\null{(#1)\null}
\usepackage{feynmp}
\makeatletter
\def\endfmffile{%
  \fmfcmd{\p@rcent\space the end.^^J%
          end.^^J%
          endinput;}%
  \if@fmfio
    \immediate\closeout\@outfmf							
  \fi
  \ifnum\pdfshellescape=\@ne
    \immediate\write18{mpost \thefmffile}%
  \fi}
\makeatother
\usepackage[ddmmyyyy]{datetime}

\usepackage{mathtools}

\DeclareMathOperator{\Tr}{Tr}

\begin{document}

\newcommand{\lij}{$\lambda_{ij}$}
\newcommand{\lam}[1]{$D_{\lambda,#1#1}$}
\newcommand{\mchi}{$m_\chi$}
\newcommand{\mphi}{$m_\phi$}
\newcommand{\tet}[1]{$\theta_{#1}$}
\newcommand{\del}[1]{$\delta_{#1}$}
\newcommand{\corres}[1]{\mathrel{\widehat{=}}}

\begin{titlepage}
\begin{flushright}
TTP17-049 
\end{flushright}
\vskip1.2cm

\begin{center}

{\LARGE \bf \boldmath Flavoured Dark Matter Moving Left}
\vskip1.0cm
{\bf \large
Monika Blanke$^{a,b}$, Satrajit Das$^c$, Simon Kast$^b$}
\vskip0.3cm
 { $^a$ Institut f{\"u}r Kernphysik, Karlsruhe Institute of Technology,\\
  Hermann-von-Helmholtz-Platz 1,
  D-76344 Eggenstein-Leopoldshafen, Germany}\vspace{1mm}\\
 { $^b$ Institut f{\"ur Theoretische Teilchenphysik},
  Karlsruhe Institute of Technology, \\
Engesserstra\ss e 7,
  D-76128 Karlsruhe, Germany}\vspace{1mm}\\
 { $^c$ Physics Department, Indian Institute of Technology Bombay,\\
 Powai, Mumbai, Maharashtra 400076, India}

\vskip0.51cm


\vskip0.35cm

{\large\bf Abstract\\[10pt]} \parbox[t]{.9\textwidth}{
We investigate the phenomenology of a simplified model of flavoured Dark Matter (DM), with a dark fermionic flavour triplet coupling to the left-handed $SU(2)_L$ quark doublets via a scalar mediator. The DM-quark coupling matrix is assumed to constitute the only new source of flavour and CP violation, following the hypothesis of Dark Minimal Flavour Violation. We analyse the constraints from LHC searches, from meson mixing data in the $K$, $D$, and $B_{d,s}$ meson systems, from thermal DM freeze-out, and from direct detection experiments. Our combined analysis shows that while the experimental constraints are similar to the DMFV models with DM coupling to right-handed quarks, the multitude of couplings between DM and the SM quark sector resulting from the $SU(2)_L$ structure implies a richer phenomenology and significantly alters the resulting impact on the viable parameter space.
}

\end{center}
\end{titlepage}




\setcounter{tocdepth}{2}


\section{Introduction}

Despite their strong theoretical motivation, until today no new particles {beyond the SM Higgs boson} have been discovered at the LHC, and the Standard Model (SM) of particle physics so far has passed all laboratory tests with flying colours (with a few glaring exceptions in the flavour sector). Yet, astrophysical and cosmological observations provide overwhelming evidence for the presence of a significant amount of Dark Matter (DM) in the Universe. The observed DM relic abundance can not be accounted for within the SM, and its particle physics origin remains a mystery.

One of the most popular DM candidates is the so-called {WIMP \cite{Arcadi:2017kky}} -- a weakly interacting massive particle which offers the intriguing possibility to link DM to the origin of electroweak symmetry breaking, predicting new particles around the TeV scale. Recently however, the non-observation of such particles in neither LHC searches nor direct DM detection experiments has challenged the simplest WIMP models and led the community to explore alternative DM models. 

A neat possibility to reconcile the WIMP paradigm with the experimental constraints is provided by the introduction of a flavour structure in the dark sector. Flavour non-universal couplings to the SM matter can efficiently suppress the DM interactions with the first SM generation, most relevant for direct detection and collider experiments. In addition, flavour symmetries have been proven to provide a natural stabilisation mechanism for DM and thereby avoid the ad hoc introduction of a discrete symmetry \cite{Batell:2011tc,DMFVPrimer}. Various simplified models of flavoured DM have been studied in the literature \cite{Kilic:2015vka,Agrawal:2011ze,Cheung:2011zza,Kile:2011mn,Batell:2011tc,Kamenik:2011nb,Kumar:2013hfa,Chang:2013oia,Kile:2013ola,Bai:2013iqa,Batell:2013zwa,Agrawal:2014ufa,Agrawal:2014una,Gomez:2014lva,DMFVPrimer,Hamze:2014wca,Lee:2014rba,Kile:2014jea,Agrawal:2015kje,Lopez-Honorez:2013wla,Blanke:2017tnb,Jubb:2017rhm,Chen:2015jkt}. 

While most of these studies imposed Minimal Flavour Violation (MFV) \cite{Buras:2000dm,D'Ambrosio:2002ex,Buras:2003jf} in order to avoid constraints from flavour precision experiments, more recently also models with a non-minimal flavour structure have been investigated \cite{DMFVPrimer,Blanke:2017tnb,Jubb:2017rhm,Chen:2015jkt}.  In \cite{DMFVPrimer} the concept of Dark Minimal Flavour Violation (DMFV) has been put forward, which assumes the coupling of the DM flavour triplet with the SM flavour triplets to constitute the only new source of flavour and CP violation. DMFV models coupling the dark sector to the right-handed down quarks, the right-handed up quarks and the right-handed charged leptons have been studied, with the outcome that the introduction of a new flavour structure significantly enriches the phenomenology of the respective models.

In the present paper we continue our expedition into the DMFV model space and investigate the possibility of a new flavour violating coupling of the DM to left-handed fermions {for the first time}. We introduce a simplified DMFV model in which the DM flavour triplet couples to the left-handed quark doublets of the SM in a non-MFV manner. The study naturally follows our previous analyses of DMFV models with couplings to the right-handed down \cite{DMFVPrimer} and up \cite{Blanke:2017tnb} quarks, respectively. It turns out that the present model combines the phenomenological implications of the scenarios studied previously in a non-trivial manner, yielding {interesting results} for the allowed parameter space of the model.

Our paper is organised as follows. In \autoref{sec:model} we introduce the structure of the simplified model {that we study} and discuss its basic features. The most important LHC constraints on the parameter space of the model are analysed in \autoref{sec:LHC}. Then in \autoref{sec:flavour}, \autoref{sec:relic}, and \autoref{sec:direct} we study in turn the constraints from precision flavour observables, from the observed relic abundance, and from direct detection experiments. In \autoref{sec:combined} we present the results of {a} combined analysis of all constraints. We conclude in \autoref{sec:summary} with a summary of our results and a brief outlook.

\section{Coupling Flavoured Dark Matter to Left-Handed Quarks}
\label{sec:model}

In this paper we study a simplified model of flavoured DM beyond the simplifying, but very restrictive assumption of Minimal Flavour Violation (MFV). The DM flavour triplet $\chi$ is introduced, in analogy to the SM matter {below the electroweak scale}, as Dirac fermion, and it interacts with the SM quark via a scalar mediator $\Phi$. While in our previous studies \cite{DMFVPrimer,Blanke:2017tnb}, the DM was assumed to couple to the right-handed down or up quarks, respectively, in the present paper we investigate its interaction with the left-handed $SU(2)_L$ quark doublets:
\begin{eqnarray}\label{NPinteractionSU2}
\mathcal{L}_{\text{int}}& = -\lambda'_{ij}\bar{q}'_{Li}\chi_{j}\Phi+h.c..
\end{eqnarray}
Here, {${q}'_{Li}=({u}'_{Li}, {d}'_{Li})^\top$} labels the left-handed SM $SU(2)_L$ quark doublet in the flavour basis, and $\lambda'_{ij}$ denotes the NP coupling between the DM flavour $\chi_j$, the mediator $\Phi$ and the quark-doublet $q'_i$ in the flavour basis. 
Just as in the previously studied DMFV models, we choose the DM flavour triplet $\chi$ to be a gauge singlet. For convenience we label its components the same way as in the up quark DMFV model \cite{Blanke:2017tnb}: 
{\begin{equation}
\chi_{j}=(\chi_u,\chi_c,\chi_t)^\top\,.
\end{equation}
Throughout our analysis we assume the top flavoured state $\chi_t$ to be the lightest of the three flavours, so that $\chi_t$ forms the entire observed DM. We will see in \autoref{sec:direct} that this case is indeed phenomenologically preferred.}
 Finally, the scalar mediator $\Phi$ transforms as weak doublet, $\Phi=(\phi_u, \phi_d)^\top$, and carries the QCD colour and hypercharge of the SM quark doublet.

Following the hypothesis of Dark Minimal Flavour Violation (DMFV) proposed in \cite{DMFVPrimer}, the DM coupling $\lambda'$ constitutes the only new source of flavour and CP violation. 
This assumption limits the number of free parameters and guarantees a stable DM candidate, in analogy to {the MFV scenario in \cite{Batell:2011tc}}.

Transforming \eqref{NPinteractionSU2} to the mass eigenstate basis, without loss of generality we can decompose the coupling as
\begin{eqnarray}\label{NPinteractionSU2decomp}
\mathcal{L}_{\text{int}}& = - (\bar{u}_{Li}\lambda_{ij}\chi_j\phi_u + h.c) - (\bar{d}_{Li}\tilde{\lambda}_{ij}\chi_j\phi_d + h.c.),
\end{eqnarray}
with {${u}_{Li}$ and ${d}_{Li}$} denoting the quarks in the mass eigenstate basis. In analogy to \cite{DMFVPrimer,Blanke:2017tnb}, we parametrise $\lambda$ by
\begin{equation}\label{eq:lambda}
 \lambda=U_{\lambda}D_{\lambda}
\end{equation}
with a diagonal matrix $D_{\lambda}$ with real and positive entries, and $U_{\lambda}$ consisting of three unitary matrices carrying a mixing angle {$\theta_{ij}$ and a phase $\delta_{ij}$ ($ij=12,13,23$)} each \cite{Blanke:2006xr}:
\begin{eqnarray}
D_{\lambda}&\hspace*{-1ex} = \hspace*{-1ex}&\text{diag}(D_{\lambda,11},D_{\lambda,22},D_{\lambda,33})\,,\qquad D_{\lambda,ii}>0\,, \vspace{1mm}\\
U_{\lambda}&\hspace*{-1ex} = \hspace*{-1ex}& U^{\lambda}_{23}U^{\lambda}_{13}U^{\lambda}_{12} \\[7pt]
	   &\hspace*{-1ex} = \hspace*{-1ex}& \left( \begin{matrix} 1&0&0\\ 0&c_{23}&s_{23}e^{-i\delta_{23}} \\ 0&-s_{23}e^{i\delta_{23}}&c_{23} \end{matrix} \right)
	   \left( \begin{matrix} c_{13}&0&s_{13}e^{-i\delta_{13}} \\ 0&1&0 \\ -s_{13}e^{i\delta_{13}}&0&c_{13} \end{matrix} \right)
	   \left( \begin{matrix} c_{12}&s_{12}e^{-i\delta_{12}}&0 \\ -s_{12}e^{i\delta_{12}}&c_{12}&0 \\0&0&1 \end{matrix} \right)\,.\nonumber           
\end{eqnarray}
Here $c_{ij}=\cos \theta_{ij}$ and $s_{ij}=\sin \theta_{ij}$.
The CKM misalignment between the left-handed up and down quarks then leads to
\begin{equation}\label{eq:lambda-tilde}
\tilde{\lambda} = V^\dag_{\text{CKM}}\lambda\,,
\end{equation}
i.\,e.\ $\lambda$ and {$\tilde\lambda$} are related to each other by the (small) CKM mixing.
Throughout this paper, for the numerical values of the CKM parameters we use the latest new physics fit results of the UTfit Collaboration \cite{Bona:2016bvr}. The new physics fit provides the best determination of the CKM matrix {in the} presence of NP contributions to FCNC processes.

We can see from the interaction Lagrangian \eqref{NPinteractionSU2decomp} that, barring the different chirality of the coupling, the quark doublet DMFV model in some sense combines the previously studied {right-handed DMFV models \cite{DMFVPrimer,Blanke:2017tnb}}. Yet its phenomenology is not a mere superposition of the effects previously identified, since the couplings in the up and down sectors are connected by \eqref{eq:lambda-tilde}. We hence can expect an interesting and non-trivial interplay of the constraints found on those models.

The mass terms {for the new particles} are again realized following the principles of DMFV:
\begin{eqnarray}\label{NPmassSU2}
\mathcal{L}_{\text{mass}}& = - m_{\chi}\bar{\chi}\chi - m_{\phi}^2\Phi^{\dagger}\Phi\,.\,
\end{eqnarray}
{i.\,e.\ at this level the different flavours of $\chi$ are degenerate.} 
Note that electroweak symmetry breaking will create a small mass splitting between $\phi_u$ and $\phi_d$ of the order of $v^2/m_\phi$. As $m_\phi \gg v$, we can safely neglect this splitting in what follows.

The different flavours of $\chi$ receive a sizable {mass splitting} which can be parametrised in terms of the usual DMFV expansion \cite{DMFVPrimer,Blanke:2017tnb}:
\begin{eqnarray} \label{DMmasssplitSU2}
m_{\chi,ij}=m_{\chi}\left(\mathds{1}+\eta_1\lambda^{\dagger}\lambda+\eta_2\tilde{\lambda}^{\dagger}\tilde{\lambda}+\mathcal{O}(\lambda^4)\right)_{ij}=m_{\chi}\left(1+\eta(D_{\lambda,ii})^2+\mathcal{O}(\lambda^4)\right)\delta_{ij}\,.
\end{eqnarray}
Note that due to its unitarity $V_{\text{CKM}}$ drops out of the above formula. In our simplified model we incorporate $\eta=\eta_1+\eta_2$ as a free parameter.


{In summary this leaves our model with twelve free parameters, which are the bare mass parameters $m_{\phi}$ and $m_{\chi}$, the nine parameters of the coupling matrix $\lambda$ (three diagonal ``couplings'' $D_{\lambda,ii}$, three mixing angles $\theta_{ij}$, and three phases $\delta_{ij}$) as well as the  parameter $\eta$, governing the mass splitting of the DM flavours. If not mentioned otherwise in the text or figures, all these parameters are scanned over in our analysis. One exception is the parameter $\eta$, which is set to a fixed benchmark value depending on the studied freeze-out scenario.}

\section{LHC Constraints}
\label{sec:LHC}

In this section we investigate the constraints that NP searches at the LHC place on the parameter space of our model.
As discussed in \cite{Papucci:2014iwa} and confirmed in \cite{DMFVPrimer} for the DMFV model with coupling to down quarks, the strongest bounds on DM models with coloured $t$-channel mediator typically originate from mediator pair production and subsequent decay. This holds true for the quark doublet DMFV model as well. Due to the $SU(2)_L$ charge of the mediator, in the present case both $\phi_u$ and $\phi_d$ production contribute to the total cross-section. Recall that $\phi_u$ couples  DM to the left-handed up quarks, while $\phi_d$ couples DM to the left-handed down quarks. 

We identify the following new features compared to the LHC phenomenology of the  up quark \cite{Blanke:2017tnb} and down quark \cite{DMFVPrimer} DMFV model, respectively:
\begin{itemize}
 \item \textbf{Production}: Both $\phi_u^\dagger\phi_u$ and $\phi_d^\dagger\phi_d$ pairs are produced via QCD interactions, and via tree level exchange of the DM triplet $\chi$ in the $t$-channel. In addition, $\chi$ exchange generates the mixed states $\phi_u^\dagger \phi_d$ and $\phi_d^\dagger \phi_u$, see \autoref{phiuphidproduction}. Thus, pair production of the scalar mediator is significantly enhancend with respect to the previously studied DMFV models.
\begin{figure}[ht!]
\centering
\begin{fmffile}{Z-phiuphidprod}
  \centering
  \quad\quad\quad\begin{fmfgraph*}(30,25)
    \fmfbottom{i1,o1}
    \fmftop{i2,o2}
    
    \fmflabel{$u_{i}$}{i2}
    \fmflabel{$\bar{d_{k}}$}{i1}
    \fmflabel{$\phi_u^\dagger$}{o2}
    \fmflabel{$\phi_d$}{o1}
    \fmf{fermion}{v1,i1}
    \fmf{fermion}{i2,v2}
    \fmf{dashes}{o1,v1}
    \fmf{dashes}{v2,o2}
    \fmf{fermion,tension=1,label=$\chi_{j}$}{v2,v1}
  \end{fmfgraph*}
\end{fmffile}
\caption{Relevant production channel for $\phi_u \phi_d$ mixed state.}
\label{phiuphidproduction}
\end{figure}
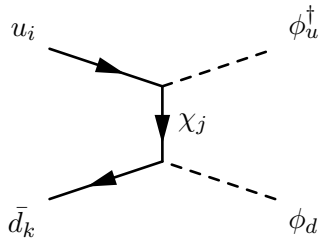
 \item \textbf{Decay}: The interaction Lagrangian \eqref{NPinteractionSU2decomp} implies that $\phi_u$  decays into a DM flavour $\chi$ plus a member of the up quark flavour triplet, while $\phi_d$ decays into $\chi$ plus a member of the down quark flavour triplet. Hence, we need to consider the following possible final states for which dedicated ATLAS and CMS searches exist:
 $t\bar{t}+\cancel{\it{E}}_{T}$ emerges from $\phi_u^\dagger \phi_u$ production and subsequent decay into the third generation quarks. The resulting limits on the parameter space are  the same as found for the up quark DMFV model \cite{Blanke:2017tnb}. Furthermore, $b\bar{b}+\cancel{\it{E}}_{T}$ is realised from the $\phi_d^\dagger \phi_d$ intermediate state and subsequent decay to third generation quarks. As the CKM mixing is safely negligible, the resulting limits are the same as in the down quark DMFV model \cite{DMFVPrimer}. The only signature to be considered is hence $\text{jets}+\cancel{\it{E}}_{T}$. This final state arises not only from $\phi_u^\dagger \phi_u$ and $\phi_d^\dagger \phi_d $ intermediate states, but also from the mixed states of $\phi_u$ and $\phi_d$ -- each time with both mediators decaying to {quarks of the  first two generations}.
\end{itemize}

While the production cross section for  the $\text{jets}+\cancel{\it{E}}_{T}$ final state is significantly increased with respect to our previous DMFV studies \cite{DMFVPrimer,Blanke:2017tnb}, the contributions to the $b\bar{b}+\cancel{\it{E}}_{T}$ and $t\bar{t}+\cancel{\it{E}}_{T}$ final states remain unchanged and the results from \cite{DMFVPrimer,Blanke:2017tnb} hold. It hence comes as no surprise that in the quark doublet DMFV model we obtain by far the strongest constraints from the $\text{jets}+\cancel{\it{E}}_{T}$ bounds, as the latter have already been found to be dominant in the {right-handed} DMFV models, respectively. {In order to allow for a straightforward comparison with the bounds on the right-handed DMFV models, in what follows we restrict ourselves to the constraints obtained from run 1 of the LHC. We note that with the rapidly increasing integrated luminosity at LHC run 2, the lower bound on the mediator mass will become more stringent.}

\begin{figure}[t]
\centering
 \begin{subfigure}[t]{0.49\textwidth}
  \centering
  \includegraphics[width=\linewidth]{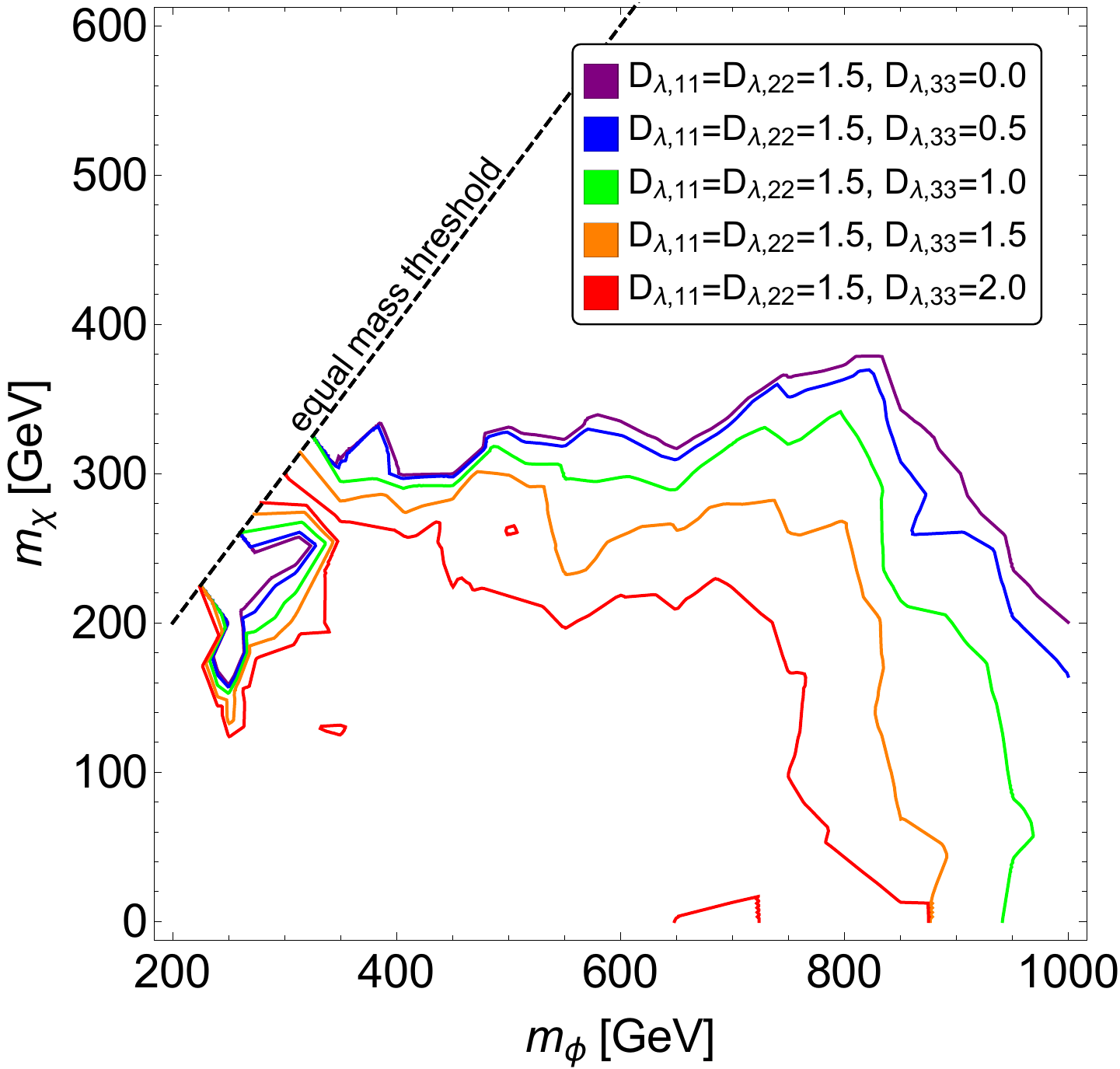}
 \end{subfigure} 
\begin{subfigure}[t]{0.49\textwidth}
  \centering
  \includegraphics[width=\linewidth]{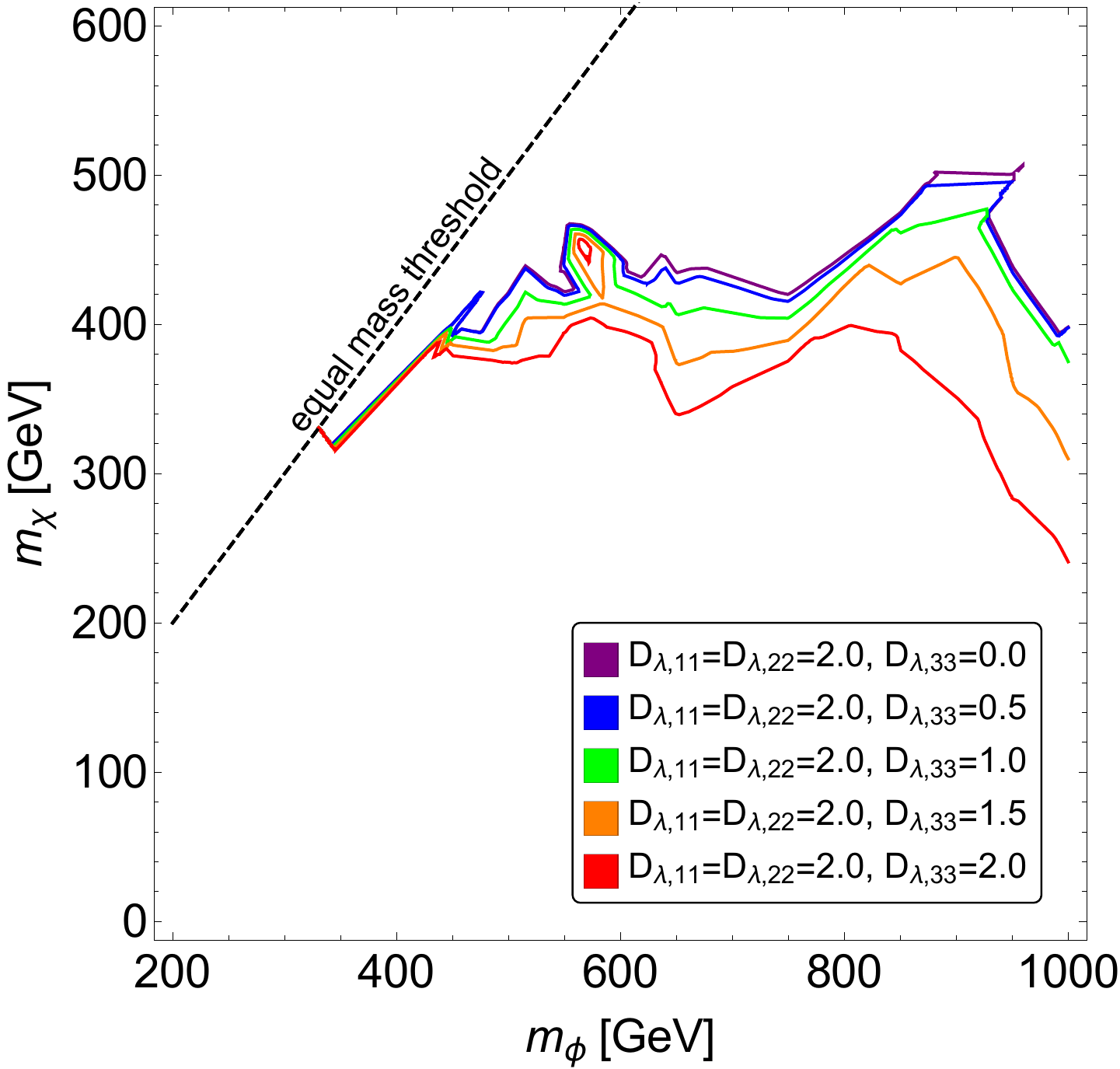}
\end{subfigure} 

\caption{Constraints on the $\text{jets}+\cancel{\it{E}}_{T}$ final state from 8\,TeV LHC run 1 data, obtained from the cross-section limits in \cite{Aad:2014wea}. Left: 95\% C.L. exclusion contours of the quark-doublet DMFV model for various strengths of third generation couplings, with fixed $D_{\lambda,11}=D_{\lambda,22}$\,=\,1.5. Right: 95\% C.L. exclusion contours of the quark-doublet DMFV model for various strengths of third generation couplings, with fixed $D_{\lambda,11}=D_{\lambda,22}$\,=\,2.0, {and the mixing angles $\theta_{ij}$ set to zero. Note that, as discussed in detail in \cite{Blanke:2017tnb}, the latter do not affect the identified safe parameter space.}\label{jj_couplings_influenceSU2}}
\end{figure}

For the phenomenological analysis we employ the same assumptions, simplifications and strategies as discussed in detail in \cite{DMFVPrimer}. For the sake of brevity we only show the exclusion bounds from the $\text{jets}+\cancel{\it{E}}_{T}$ final state, see \autoref{jj_couplings_influenceSU2}. We observe that the constraints for equivalent coupling strengths exclude a significantly larger region of the \mphi\,-\,$m_{\chi_t}$ plane than in the up quark DMFV model. We do not show the exclusion bounds for mediator masses larger than 1\,TeV, since in this range numerical uncertainties from extrapolation of the exclusion data given in \cite{Aad:2014wea} dominate.

From the results shown in \autoref{jj_couplings_influenceSU2} we learn that for mediator masses as low as 
\begin{equation}\label{eq:mphi}
m_\phi=850\,\text{GeV}
\end{equation}
 the DM couplings have to be restricted to 
\begin{equation}\label{eq:Dmax}
D_{\lambda,11},D_{\lambda,22} \le D_{\lambda,33}\le 1.5
\end{equation}
in order to satisfy the LHC constraints. 
Note that the upper bound on the couplings for a given mediator mass  is significantly lower in the quark doublet model than in the {right-handed} DMFV models \cite{DMFVPrimer,Blanke:2017tnb}. 
{In what follows we restrict the parameter space of the model to the ``collider-safe'' region defined by \eqref{eq:mphi} and \eqref{eq:Dmax}.}

Before moving on let us point out that 
{the quark doublet DMFV model predicts various flavour violating signatures at the LHC. As in the right-handed DMFV model coupling to up quarks \cite{Blanke:2017tnb}, the final state $tj+\cancel{\it{E}}_{T}$ is produced with significant rate, even in the limit of no flavour mixing, $\theta_{ij}=0$.\footnote{This is to be contrasted with the case of supersymmetric models, where such signatures are generated only in the presence of large flavour mixing \cite{Hurth:2009ke,Blanke:2013zxo,Agrawal:2013kha,Arana-Catania:2014ooa,Backovic:2015rwa,Blanke:2015ulx}.} Additionally, the present model gives rise to the smoking gun signature $tb+\cancel{\it{E}}_{T}$ which becomes relevant for sizable DM couplings $D_{\lambda,ii}$. This signature arises from the mixed $\phi_u \phi_d$ production with both mediators decaying to the third quark generation and is hence present only if DM couples to the SM quark doublets. A measurement of the $tb+\cancel{\it{E}}_{T}$ cross section {would therefore provide} a direct measure of the DM coupling strength, as the QCD production channels do not contribute in this case. Note that such final state is not generated in supersymmetric models with significant rate. While these signatures in principle have some impact on single-top studies already preformed at the LHC, we expect that a dedicated search including a cut on $\cancel{\it{E}}_{T}$ would have a significantly improved reach, due to the much reduced SM background.}

\section{Flavour Constraints}
\label{sec:flavour}


Having restricted the parameter space of our model to a region that complies with present LHC searches, we now move on to the study of precision flavour observables. In analogy to the {right-handed} DMFV models \cite{DMFVPrimer,Blanke:2017tnb}, the most stringent constraints on the quark doublet model stem from meson antimeson mixing observables, while the effects on rare decays are negligible.

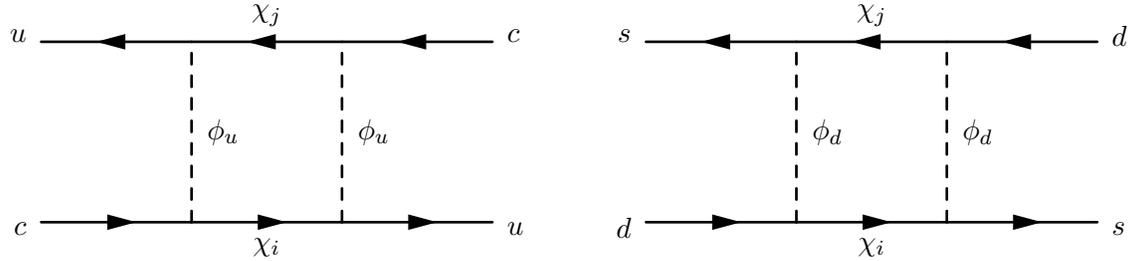
\begin{figure}[b]
\centering
\begin{fmffile}{Z-boxDmesons}
  \begin{fmfgraph*}(60,30)
    \fmfbottom{i1,d1,o1}
    \fmftop{i2,d2,o2}
    
    \fmflabel{$c$}{i1}
    \fmflabel{$u$}{i2}
    \fmflabel{$u$}{o1}
    \fmflabel{$c$}{o2}
    \fmf{fermion}{i1,v1}
    \fmf{fermion}{v2,o1}
    \fmf{fermion,label=$\chi_{i}$}{v1,v2}
    \fmf{fermion}{o2,v4}
    \fmf{fermion}{v3,i2}
    \fmf{fermion,label=$\chi_{j}$}{v4,v3}
    \fmf{dashes,tension=0,label=$\phi_u$}{v1,v3}
    \fmf{dashes,tension=0,label=$\phi_u$}{v2,v4}
  \end{fmfgraph*}
\end{fmffile}
\qquad\qquad
\begin{fmffile}{Z-boxKmesons}
  \begin{fmfgraph*}(60,30)
    \fmfbottom{i1,d1,o1}
    \fmftop{i2,d2,o2}
    
    \fmflabel{$d$}{i1}
    \fmflabel{$s$}{i2}
    \fmflabel{$s$}{o1}
    \fmflabel{$d$}{o2}
    \fmf{fermion}{i1,v1}
    \fmf{fermion}{v2,o1}
    \fmf{fermion,label=$\chi_{i}$}{v1,v2}
    \fmf{fermion}{o2,v4}
    \fmf{fermion}{v3,i2}
    \fmf{fermion,label=$\chi_{j}$}{v4,v3}
    \fmf{dashes,tension=0,label=$\phi_d$}{v1,v3}
    \fmf{dashes,tension=0,label=$\phi_d$}{v2,v4}
  \end{fmfgraph*}
\end{fmffile}
\caption{New one loop contribution to $D^0-\bar D^0$ (left) and $K^0-\bar K^0$ mixing (right). Analogous contributions to $B_d^0-\bar B^0_d$ and $B_s^0-\bar B_s^0$ mixings exist.}
\label{mesonmix}
\end{figure}

Since, due to the $SU(2)_L$ structure in the DM-quark coupling, the NP particles in this model couple to all six quark flavours, we have to consider constraints from $D^0-\bar D^0$ mixing as well as from $K^0-\bar K^0$ mixing and $B_{d,s}^0-\bar B_{d,s}^0$ mixing. The relevant one loop box diagrams for $D^0-\bar D^0$ and $K^0-\bar K^0$ mixing are shown in \autoref{mesonmix}, the contributions to $B^0_{d,s}-\bar B^0_{d,s}$ mixing are analogous.

Calculating the contributions to the dispersive part of the off-diagonal mixing amplitude, we find in analogy to the results of \cite{DMFVPrimer,Blanke:2017tnb}
\begin{eqnarray}
\label{m12Dnew}
M_{12}^{D,\text{new}} & = & \frac{1}{384\pi^{2}m_{\phi}^{2}} \,\eta_{D}\, m_{D}\,f_{D}^{2}\,{{B}_{D}} \left(({\lambda}{\lambda}^\dagger)_{cu}\right)^2\cdot L(x_{i},x_{j})\,,\\\label{m12Knew}
M_{12}^{K,\text{new}} & = & \frac{1}{384\pi^{2}m_{\phi}^{2}} \,\eta_{2}\, m_{K}\,f_{K}^{2}\,\hat{B}_{K} \left((\tilde{\lambda}\tilde{\lambda}^\dagger)_{sd}\right)^2\cdot L(x_{i},x_{j})\,,\\
\label{m12Bnew}
M_{12}^{B_q,\text{new}} & = & \frac{1}{384\pi^{2}m_{\phi}^{2}} \,\eta_{B}\, m_{B_q}\,f_{B_q}^{2}\,\hat{B}_{B_q} \left((\tilde{\lambda}\tilde{\lambda}^\dagger)_{bq}\right)^2\cdot L(x_{i},x_{j}) \qquad(q=d,s) \,.
\end{eqnarray}
{The loop function $L(x_{i},x_{j})$, with $x_i = m_{\chi_i}^2/m_\phi^2$, has been calculated in \cite{DMFVPrimer}, and can be found together with the {QCD input parameters} in the appendix. 
 The mass splitting between the DM flavours constitutes a small correction, and we neglect it in what follows.}

Note that while  \eqref{m12Dnew} includes the coupling $\lambda$, \eqref{m12Knew} and \eqref{m12Bnew} include the coupling $\tilde{\lambda}$ and hence are sensitive to the CKM mixing of the SM. Using the parametrisation of $\lambda$ and $\tilde\lambda$ in \eqref{eq:lambda}, we observe that the unitary matrices $V_\text{CKM}$ and $U_\lambda$ cancel in the expressions \eqref{m12Dnew}--\eqref{m12Bnew} in the degeneracy limit for the couplings $D_{\lambda,ii}$. The size of the NP contributions to the neutral meson mixing observables is hence suppressed by the splittings $\Delta_{ij} = |D_{\lambda,ii}-D_{\lambda,jj}|$. 

Defining
\begin{eqnarray}
 M^{B_q}_{12}=C_{B_q}e^{2i{\phi_{B_q}}}M^{{B_q},\text{SM}}_{12}\quad(q=d,s)
\end{eqnarray}
as well as
\begin{eqnarray}
 \text{Re}M^K_{12}&=&C_{\Delta{M_K}}\text{Re}M^{K,\text{SM}}_{12}\,, \\
 \text{Im}M^K_{12}&=&C_{\epsilon_K}\text{Im}M^{K,\text{SM}}_{12}\,,
\end{eqnarray}
we can compare our predictions for $C_{\Delta{M_K}}$, $C_{\epsilon_K}$, $C_{B_d}$, $\phi_{B_d}$, $C_{B_s}$ and $\phi_{B_s}$ to the allowed ranges collected in \autoref{FlavourConstraintsTableSU2}. In the case of $D^0-\bar D^0$ mixing, we make the conservative assumption that the long distance dominated SM contribution to
\begin{equation}
x_{12}^{D} = \frac{2 |M_{12}^D|}{\Gamma_D} = 2 |M_{12}^D| \tau_{D^0}
\end{equation}
is at most $\pm 3\%$. With 
\begin{equation}
\Phi^D_{12} = \arg M_{12}^D
\end{equation}
in the standard phase conventions, the bounds in \autoref{FlavourConstraintsTableSU2} can be directly applied to our model. The necessary input parameters are collected in the appendix.

\begin{table}[t]
\centering
       \begin{tabular}{| c | l || c | l |}
 \hline
 \rule{0pt}{2ex}
 $C_{\Delta{M_K}}$ & $1.10\pm0.44$ & $C_{B_d}$ & $\in[0.81, 1.28]$\tabularnewline
 $C_{\epsilon_K}$ & $\in[0.83, 1.28]$ & $\phi_{B_d}$ & $\in[-5.2^\circ, 1.5^\circ]$\tabularnewline
 \hline
 \rule{0pt}{3ex}
  $x^D_{12}$ & $\in[0.10\%,0.67\%]$ & $C_{B_s}$ & $\in[0.899, 1.252]$\tabularnewline
  $\Phi^D_{12}$  & $\in[-5.3^\circ,4.4^\circ]$ & $\phi_{B_s}$ & $\in[-1.848^\circ, 1.959^\circ]$\tabularnewline
 \hline
       \end{tabular}
\caption{Constraints on the $K^0-\bar K^0$, $D^0-\bar D^0$ and $B_{d,s}^0-\bar B_{d,s}^0$ systems \cite{Bona:2005eu,Bona:2007vi,Amhis:2016xyh}.}
\label{FlavourConstraintsTableSU2}
\end{table}

\begin{figure}
  \centering
  \includegraphics[width=.52\linewidth]{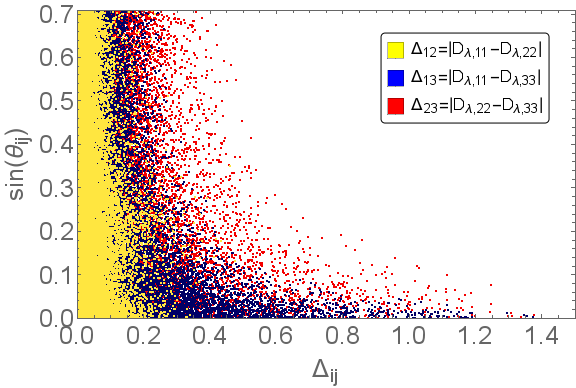}
\caption{Viable ranges for the flavour mixing angles \tet{ij} of the quark doublet DMFV model in dependence of the splittings between couplings \lam{i} and \lam{j}, for mediator mass \mphi\,=\,850\,GeV and DM mass \mchi\,=\,250\,GeV. {Different colours correspond to the different mixing angles $\theta_{ij}$ and splittings $|D_{\lambda,ii}-D_{\lambda,jj}|=\Delta_{ij}$, with each colour representing one pair of indices:
$ij=12$ in yellow, $ij=13$ in blue, $ij=23$ in red.}}
\label{FC_valid_areaSU2}
\end{figure}

Imposing the constraints from meson antimeson mixing in this way, we obtain the viable ranges for the flavour mixing angles $\theta_{ij}$ as function of the respective splitting $|D_{\lambda,ii}-D_{\lambda,jj}|=\Delta_{ij}$, see \autoref{FC_valid_areaSU2}.
We observe that the structure of the coupling matrix $\lambda$ is significantly more constrained than in the case of the up quark DMFV model \cite{Blanke:2017tnb}, due to the {additional} strong constraints from {$K^0-\bar K^0$ and $B_{d,s}^0-\bar B_{d,s}^0$ mixing} data. In particular we find significant effects on $\theta_{13}$ and $\theta_{23}$ as well. The constraints on $\theta_{12}$ remain the most severe, and in fact the combination of $K^0-\bar K^0$ and $D^0-\bar D^0$ mixing constraints rejects any splitting $\Delta_{12}$ larger than {approximately 0.3}.  This is a direct consequence of the CKM misalignment between the left-handed up and down quarks with a rather large $|V_{us}| \approx 0.225$ -- a sufficient alignment of the NP sector with both the up and down sectors cannot be achieved for large splittings. 

Comparing the allowed ranges in \autoref{FC_valid_areaSU2} with those obtained in Figure 5 of \cite{DMFVPrimer}, it appears as if our upper bounds on the flavour mixing angles are less severe -- apart from the cut for a too large $\Delta_{12}$. This seems unexpected, since the quark doublet model is constrained by the same bounds as the down quark DMFV model plus  the limits from $D^0-\bar D^0$ mixing. The reason for this difference is the different approaches in scanning the couplings \lam{i}. While in \cite{DMFVPrimer} in the flavour analysis only the flavour violating parameters were varied and the trace of $D_\lambda$ was fixed to
$\Tr D_\lambda = 3$,
in the present study we vary all couplings $D_{\lambda,ii}$ independently between 0 and 1.5, thereby allowing for
$0\le \Tr D_\lambda \le 4.5$. It is hence not surprising that larger viable parameter ranges are found in \autoref{FC_valid_areaSU2}.

\section{Relic Abundance Constraint}
\label{sec:relic}

In this section we investigate the constraints on the quark doublet DMFV model imposed by the observed relic abundance, assuming the DM to be the relic of a thermal freeze-out. In this model the DM flavours can annihilate to all six quark flavours of the SM. The number of possible final states is thus larger than in the right-handed DMFV models \cite{DMFVPrimer,Blanke:2017tnb}.
The effective annihilation cross section is then given as the sum of the annihilation cross sections into up and down final states:
\begin{equation}\label{SU2_annihil}
\braket{\sigma v}_\text{eff}=\braket{\sigma v}_\text{eff}^u+\braket{\sigma v}_\text{eff}^d\,.
\end{equation}
In order to reproduce the observed relic abundance, the effective annihilation cross section has to satisfy \cite{Steigman:2012nb,Ade:2015xua}
\begin{equation}\label{eq:sigmav-required}
\braket{\sigma v}_\text{eff}\simeq 2.0\cdot 10^{-26} \,\text{cm}^3/\text{s}\,.
\end{equation}


As discussed in detail in \cite{DMFVPrimer,Blanke:2017tnb}, different freeze-out scenarios are possible depending on the mass splitting between the DM flavours. If the mass splitting between the DM $\chi_t$ and the heavier flavours $\chi_u,\chi_c$ is {small, $ < 1\%$,} then the lifetime of the heavier states is long enough that they fully contribute to the freeze-out process. This scenario is called {\it quasi-degenerate freeze-out (QDF)}.

In the QDF scenario the contribution from up quark final states to the effective annihilation cross section reads \cite{Blanke:2017tnb}
\begin{equation} \label{coancs}
{\braket{\sigma v}^u_\text{eff}}=\frac{1}{18}\cdot\frac{3}{32\pi}\cdot\frac{1}{4}\sum_{i,j=1}^3\sum_{k,l=u,c,t}|\lambda_{ki}|^2|\lambda_{lj}|^2\,\frac{\sqrt{\left(4m_{\chi}^{2}-(m_{k}-m_{l})^{2}\right)\left(4m_{\chi}^{2}-(m_{k}+m_{l})^{2}\right)}}{\left(m_{\phi}^{2}+m_{\chi}^{2}-\frac{m_{k}^{2}}{2}-\frac{m_{l}^{2}}{2}\right)^{2}},
\end{equation}
where an averaging factor $1/9$ for the $3\times 3$ different initial states has been included. For DM masses below $m_t$ the $t\bar t$ final state becomes kinematically inacessible and has to be excluded. Similarly for $m_\chi<m_t/2$ all final states containing a top quark become inaccessible.
In the down quark contribution the quark masses can be neglected and the expression simplifies to \cite{DMFVPrimer} 
\begin{equation}\label{SU2_down_annihil}
\braket{\sigma v}_\text{eff}^d=\frac{1}{18}\cdot\frac{3}{32\pi}\sum_{i,j=1}^3\sum_{k,l=d,s,b}\frac{|\tilde{\lambda}_{ki}|^2|\tilde{\lambda}_{lj}|^2m^2_\chi}{\left(m^2_\chi+m^2_\phi\right)^2}.
\end{equation}

If on the other hand the splitting of the heavier states with respect to the light stable flavour $\chi_t$ is large, then their decay is sufficiently rapid that these states are no longer present at the time of freeze-out. This scenario is referred to as {\it single flavour freeze-out (SFF)}. In this case only the $\chi_t\bar \chi_t$ initial state contributes to the effective annihilation cross section and the flavour averaging factor $1/9$ is absent.

\begin{figure}
\centering
  \includegraphics[width=0.5\linewidth]{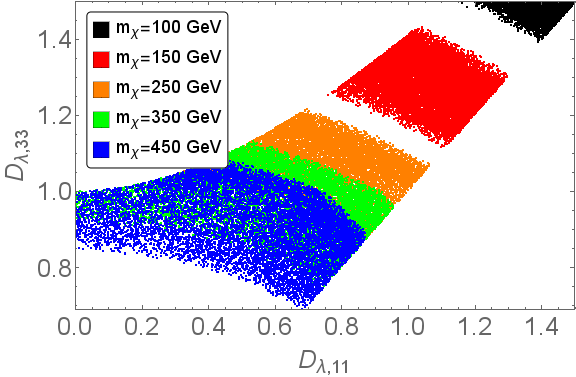}

\caption{Viable regions of parameter space of the quark-doublet DMFV model in the QDF scenario (with $m_\phi = 850\,\text{GeV}$ and $\eta=-0.01$) compatible with the relic abundance constraint, at different DM masses.}
\label{RA_valid_area_SU2}
\end{figure}

The phenomenological implications of the thermal freeze-out condition \eqref{eq:sigmav-required} are  quite similar to the ones observed for the up quark DMFV model in \cite{Blanke:2017tnb}. We therefore refer the interested reader to section 5 of that paper for a detailed discussion.
Here we restrict ourselves to pointing out the differences with respect to
the up quark model:
\begin{itemize}
 \item Due to the larger number of final states, at a fixed mediator mass \mphi\ and fixed coupling matrix $\lambda$, the relic abundance constraint \eqref{eq:sigmav-required} demands a smaller DM mass in this model, relative to the up quark DMFV model. We can observe this in \autoref{RA_valid_area_SU2}, when compared to figure 5.2 of \cite{Blanke:2017tnb}. Therefore, the obtained lower DM mass bound is similar to the one obtained in \cite{Blanke:2017tnb}, although the LHC constraints studied in \autoref{sec:LHC} force us to impose a more stringent upper bound on the DM coupling strength.
 \item Since more final states exist for the DM annihilation process, the impact of passing the threshold for the $t\bar t$ final state is less significant.
\end{itemize}

\section{Direct Detection {Constraints}}
\label{sec:direct}


Last but not least, in this section we consider the constraints from direct detection experiments that search for the scattering of DM particles off the nuclei of the target material. The current best limits on the WIMP nucleon scattering cross section are set by experiments using xenon as target material, in particular XENON1T \cite{Aprile:2017iyp}, and major improvements are expected over the coming years.

The WIMP nucleon scattering is dominated by spin-independent contributions, due to coherence effects. In the quark doublet DMFV model, in analogy to the {right-handed} DMFV models \cite{DMFVPrimer,Blanke:2017tnb}, the total spin-independent cross section reads
\begin{equation} \label{WIMPnucleonCSnatSU2}
 \sigma_{n,\text{nat-Xe}}^\text{SI} = \sum_{i=1}^{9} \rho_i \cdot \frac{\mu_n^2}{\pi A_i^2}\big|Z f_p + (A_i-Z)f_n\big|^2\,,
\end{equation}
where $\mu_n$ is the reduced mass of the WIMP nucleon system. In writing \eqref{WIMPnucleonCSnatSU2} we sum over all stable and quasi-stable xenon isotopes. Their atomic mass numbers $A_i$ and natural abundance $\rho_i$ can be found in Table 2 of \cite{Blanke:2017tnb}. 

The scattering amplitudes off protons and neutrons, $f_p$ and $f_n$,
receive contributions from the DM couplings to both up and down quarks. These contributions turn out to be the same as in the right-handed DMFV models \cite{DMFVPrimer,Blanke:2017tnb}. For completeness, their analytic expressions are listed in the appendix. The relevant contributions are:
\begin{itemize}
\item
Tree level exchange of the mediator $\phi_{u,d}$. These contributions are {positive and} proportional to the flavour mixing between the first and third generation, $\sin^2\theta_{13}$.
\item
Box diagrams yield contributions to both $f_p$ and $f_n$. These are always positive.
\item
The photon penguin only contributes to $f_p$ and is positive for {virtual} up quarks {in the loop}, but negative for {virtual} down quarks. This additional negative contribution, relative to the up quark DMFV model, opens up the possibility for more diverse cancellation patterns of the various contributions.
\item
The $Z$ penguin contribution with the top quark in the loop. As this contribution is proportional to the {squared} quark mass in the loop, the effects of the light quarks are negligible. Note that the $Z$ penguin provides the only negative contribution to $f_n$.
\end{itemize}

\begin{figure}
  \centering
  \includegraphics[width=.55\linewidth]{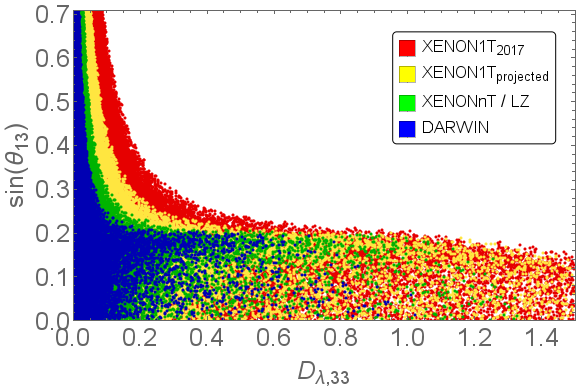}
\caption{Viable areas of parameter space of the quark doublet DMFV model, for different WIMP nucleon scattering cross section limits, for DM mass \mchi\,=\,250\,GeV and mediator mass \mphi\,=\,850\,GeV. The current exclusion bounds of XENON1T \cite{Aprile:2017iyp}, as well as the projected bounds of XENON1T \cite{Diglio:2016stt}, XENONnT \cite{Diglio:2016stt}, LUX-ZEPLIN (LZ) \cite{Akerib:2015cja} and DARWIN \cite{Aalbers:2016jon} have been imposed respectively. {Note that in this scan we did not impose any of the other constraints discussed before, as we focus on the bare influence of the direct detection bounds.}}
\label{DDExpMultiBoundsSU2}
\end{figure}

\autoref{DDExpMultiBoundsSU2} shows the allowed values of the flavour mixing angle \tet{13} in dependence of the coupling \lam3 of the DM $\chi_t$ to {the visible SM} matter. We show the allowed regions for various strength of the direct detection bounds. The current XENON1T bounds \cite{Akerib:2016vxi}, as well as the  projected bounds from XENON1T \cite{Diglio:2016stt}, XENONnT \cite{Diglio:2016stt}, LUX-ZEPLIN (LZ) \cite{Akerib:2015cja} and DARWIN \cite{Aalbers:2016jon} have been imposed respectively. 

For small coupling \lam3 we observe all \tet{13} values to be allowed. This is due to the overall suppression of the scattering cross section by (\lam3)$^4$, as discussed in detail for the up quark DMFV model in \cite{Blanke:2017tnb}. Also similar to the up quark DMFV model, for larger \lam3, an upper bound on the flavour mixing angle \tet{13} can be observed. The reason is the suppression of the tree level contributions that is required in order to  allow for a sufficient interference between the positive and negative contributions discussed above. For the same mediator mass and DM mass, we find the upper bound at a slightly lower value than in the up quark DMFV model. This shift originates in the additional tree level contributions from down quark scattering. Since an average xenon isotope contains more neutrons than protons, the tree level contribution to DM-xenon scattering from down quark scattering is larger than the one from up quark scattering. At the same time, the additional negative contribution from the photon penguin with down quark flavours in the loop are smaller in magnitude than the negative $Z$ penguin with the top quark in the loop. Hence, the tree level contributions need to be more strongly suppressed compared to the up quark DMFV model. 

This shifted balance between positive and negative contributions is also the reason why, in contrast to the up quark DMFV model, we do not find a lower bound on \tet{13} for any value of \lam3. Even if the tree level contributions are absent, the increased number of positive contributions can still be large enough to sufficiently cancel the large (in magnitude) negative $Z$ penguin and the smaller (in magnitude) negative photon penguin contribution.

In addition, we observe that this changed cancellation pattern allows for larger (compared to the up quark DMFV model) \lam3 values for the more stringent projected bounds from future xenon experiments. This is due to the fact that in this model there is both a negative contribution to the proton coupling and to the neutron coupling, suppressing both $f_n$ and $f_p$ independently. Hence a sufficient suppression of the scattering cross section is easier to achieve for the multiple xenon isotopes. Nevertheless, we still observe that more stringent bounds significantly constrain the allowed \lam3 values. 

For completeness we need to mention that, {as in the up quark DMFV model \cite{Blanke:2017tnb}}, the top flavoured DM case is still the preferred scenario in this model. The large negative $Z$ penguin contribution to $f_n$ is the most efficient way to cancel the multitude of positive contributions and hence allows for the largest viable parameter space.

\section{Combined Analysis}
\label{sec:combined}


Having studied the implications of the experimental constraints from flavour data, the observed relic density, and direct detection experiments, what remains to be done is to conduct a combined analysis, which is the subject of this section. Note that the LHC constraints are taken into account by restricting the parameter space of our model as discussed in \autoref{sec:LHC}.
 
\begin{figure}
 \begin{subfigure}{0.48\textwidth}
  \centering
  \includegraphics[width=.95\linewidth]{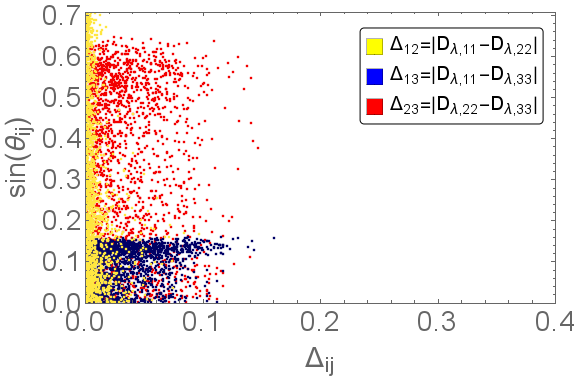}
  \caption{QDF, $m_\chi=150$\,GeV}
  \label{SU2Com-deg:sfig1}
 \end{subfigure}
 \hfill
\begin{subfigure}{0.48\textwidth}
  \centering
  \includegraphics[width=.95\linewidth]{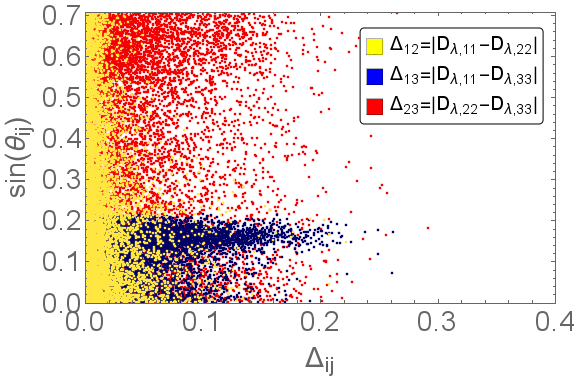}
  \caption{QDF, $m_\chi=450$\,GeV}
  \label{SU2Com-deg:sfig2}
\end{subfigure}\vspace{3mm}
\caption{{Viable} regions of parameter space of the quark doublet DMFV model, with combined flavour, relic abundance and direct detection constraints applied, for QDF scenario and different DM masses, with $m_\phi=850$\,GeV and $\eta = -0.01$. Different colours correspond to the different mixing angles $\theta_{ij}$ and splittings $|D_{\lambda,ii}-D_{\lambda,jj}|=\Delta_{ij}$, {with each colour representing one pair of indices}:
$ij=12$ in yellow, $ij=13$ in blue, $ij=23$ in red.}
\label{SU2Com-deg}
\end{figure}

For the QDF scenario, \autoref{SU2Com-deg} shows the viable parameter ranges for the coupling matrix $\lambda$ that remain after imposing the flavour, relic abundance and direct detection constraints simultaneously. We observe that the allowed splittings $|D_{\lambda,ii}-D_{\lambda,jj}|$ are very restricted due to the combination of the strong flavour constraints and the freeze-out scenario splitting conditions. The flavour mixing angle \tet{13}  is bounded from above, which is the consequence of the combination of relic abundance and direct detection constraints. Interestingly, for the largest allowed splittings $\Delta_{13}$ the mixing angle \tet{13} is required to take values close to the upper bound. The reason can be found in the direct detection constraint. A larger splitting $\Delta_{13}$ in QDF with a top-flavoured DM candidate means a smaller \lam1 compared to \lam3. Hence, the positive box diagram contributions are smaller compared to the absolute size of the negative $Z$ penguin. To still allow for a sufficient cancellation, the mixing angle \tet{13} must not be too small, allowing for a relevant tree level and not too large (in magnitude) $Z$ penguin contribution. 

\begin{figure}
  \centering
  \includegraphics[width=.55\linewidth]{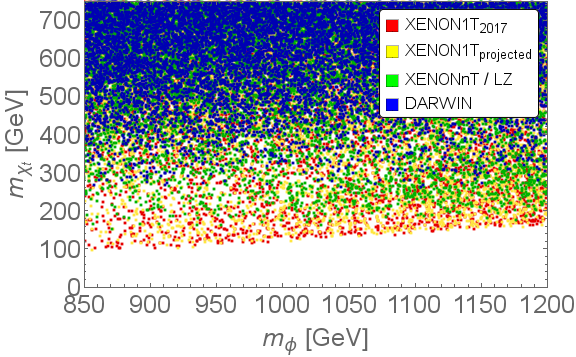}
\caption{{Viable} area in the \mphi\,-\,$m_{\chi_t}$ plane of the quark doublet DMFV model with combined flavour, relic abundance and direct detection constraints applied in the QDF scenario, for $\eta=-0.01$. We show the {viable ranges} for different strengths of direct detection bounds. The current exclusion bounds of XENON1T \cite{Aprile:2017iyp} as well as the projected bounds of XENON1T \cite{Diglio:2016stt}, XENONnT \cite{Diglio:2016stt}, LUX-ZEPLIN (LZ) \cite{Akerib:2015cja} and DARWIN \cite{Aalbers:2016jon} are applied respectively. The interference of relic abundance and direct detection effects results in a lower bound on the DM mass, increasing with more stringent direct detection bounds.}
\label{Mass-BoundsSU2}
\end{figure}

In \autoref{Mass-BoundsSU2} we show the allowed regions in the \mphi\,-\,$m_{\chi_t}$ plane in the QDF scenario for different strengths of the direct detection exclusion limits. We observe that the {current lower DM mass bound is similar to the one found in the up quark DMFV model. The reason is that the changed relic abundance phenomenology, requiring smaller couplings, counterbalances the more stringend LHC constraints. Details can be found in \autoref{sec:relic} and \autoref{sec:LHC}, respectively.}

Furthermore we find that the changed cancellation pattern in the various contributions to the WIMP nucleon scattering cross section allows for larger (compared to the up quark DMFV model) \lam3 when the projected limits from future direct detection experiments are imposed.
This (in addition to the previously explained relic abundance effect) explains why the lower DM mass bound from the combined analysis is less stringent (compared to the up quark DMFV model) for the future direct detection bounds. For the current XENON1T limit however the weaker direct detection constraints are compensated by the stronger LHC bounds, which led us to impose a lower upper limit on the couplings $D_{\lambda,ii}$ than in the up quark model.

We show no results for the SFF scenario, since the combined {bounds disfavour} this scenario for the quark doublet model.  The reason are the more stringent flavour constraints in combination with the direct detection constraints on \tet{13}. To realise the SFF scenario, a sufficiently large splitting between the coupling \lam3 and the other couplings is necessary. For a significant splitting between the couplings, the flavour constraints require small flavour mixing angles \tet{13} and \tet{23}. At the same time a large splitting between \lam3 and the other couplings results in a relatively small size of the direct detection box diagram contributions compared to the absolute size of the negative $Z$ penguin contribution. As discussed before, this requires \tet{13} to be sufficiently large, which then is in conflict with the flavour constraints. Only for {large values  $\eta\ge 0.3$} it is possible to generate a sufficient mass splitting from small splittings {$D_{\lambda,33}-D_{\lambda,ii}\ (i=1,2)$.}

\section{Summary and Outlook}
\label{sec:summary}

To conclude our analysis of the quark doublet DMFV model, we now give a summary of all observed constraints from the multitude of data. We focus mainly on the differences compared to the right-handed DMFV models \cite{DMFVPrimer,Blanke:2017tnb}. The following crucial differences have been identified:
\begin{itemize}
 \item 
The limits from LHC searches result in considerably stronger constraints, due to both additional production channels as well as the additional decay modes. To avoid constraints for the phenomenologically interesting parameter region, the couplings have to be more strictly constrained than in the right-handed DMFV models.
 \item 
Since the dark sector in the quark doublet model couples to all quark flavours of both up- and down-type, the model is affected by the constraints from meson mixing data in the $K^0-\bar K^0$, $D^0-\bar D^0$, and $B^0_{d,s}-\bar B_{d,s}^0$ systems. This results in significantly stronger bounds on the flavour mixing angles of the coupling matrix $\lambda$ than in the up quark model which only contributed to $D$ meson phenomenology. In case of \tet{12}, the constraints are also more stringent than in the down quark model. Since now $D^0-\bar D^0$  and $K^0-\bar K^0$ mixing bounds have to be satisfied simultaneously, the splitting between \lam1 and \lam2 is bounded from above, as due to the CKM misalignment small values of \tet{12} are not enough to obtain a sufficient suppression of the NP contributions.
 \item
 Due to the larger number of possible final states in DM annihilation, the required coupling strength for fixed DM and mediator masses is lower than in the up quark DMFV model. This effect {roughly} compensates the more stringent upper bound on the couplings \lam{i} from the LHC searches, so that the lower bound on the DM mass remains similar.
 \item 
The number of relevant contributions to the WIMP-nucleon scattering cross section is larger, resulting in a more {diverse} cancellation pattern in order to suppress the model prediction below the exclusion limits from direct detection experiments. In contrast to the right-handed models, we now have two negative contributions which allow to suppress both the scattering amplitudes with protons and neutrons. This not only changes the viable parameter ranges with respect to the right-handed DMFV models, but it also softens the constraints from future xenon experiments for which the isotope composition of natural xenon becomes more relevant.
\end{itemize}

The full power of the constraints described above can only be appreciated in a combined analysis. We have found the following main implications for the viable parameter space of the model:
\begin{itemize}
 \item 
The stronger flavour bounds in combination with the relic abundance and direct detection constraints require small splittings between the couplings \lam{i}. The QDF scenario is hence favoured over the SFF scenario. For the same reason, two flavour freeze-out scenarios requiring a significant splitting between \lam1 and \lam2 are ruled out.
 \item 
The combination of LHC, relic abundance and direct detection constraints result in a lower bound on the DM mass, becoming stronger with future improved direct detection cross section limits. The bound is however more relaxed than in the up quark DMFV model, thanks to the smaller couplings required by the thermal freeze-out condition.
 \item 
Keeping in mind that the lowest valid DM masses demand the largest allowed couplings, we have to conclude that the bounds from LHC searches result in the most stringent limits for a large part of the phenomenologically interesting parameter space. LHC run 2 data will hence push the exclusion limits even further. In contrast to the {right-handed} DMFV models, collider constraints prove to be the most efficient way to {test} a significant part of the parameter space of the quark doublet DMFV model, in particular reaching to increased mediator masses $m_\phi$ and significant couplings. Nevertheless, it is the combination of relic abundance and direct detection constraints that enables us to exclude low DM masses for all mediator masses, which remains an advantage over the pure collider bounds.
\end{itemize}

{Finally a brief comment is in order concerning possible signals in indirect detection experiments from DM annihilation. Annihilation in the early universe can affect reionisation, constrained by the PLANCK experiment. With the lower bound on the DM mass of at least $100\,\text{GeV}$, the current data are however not sensitive enough to probe the annihilation cross-section required for thermal freeze-out  \cite{Liu:2016cnk}. Annihilation of $b$-flavoured DM has also been shown to provide a possible explanation for the galactic center gamma-ray excess \cite{Agrawal:2014una}. However, also in this case the lower bound on the DM mass is in conflict with the required mass scale of about $50\,\text{GeV}$.
}

The analysis presented in this paper completes our exploratory study of quark flavoured DM beyond MFV, having {investigated} in turn the phenomenology of coupling to the right-handed down quarks \cite{DMFVPrimer}, right-handed up quarks \cite{Blanke:2017tnb}, and the left-handed quark doublets.  {Throughout this journey,} Dark Minimal Flavour Violation has proven to serve as a simple yet efficient concept to allow for new flavour and CP violating effects in the coupling to the dark sector, thereby significantly altering the predicted phenomenology. While restricting ourselves to the study of simplified models is phenomenologically justified, it would {certainly also} be interesting to investigate possible embeddings of the analysed setups into more complete theories.

\paragraph{Acknowledgements}

We are grateful to the Karlsruhe House of Young Scientists (KHYS) for supporting S.\,D.'s
 internship at the KIT. S.\,D.\ acknowledges the hospitality of the KIT Institute of Theoretical Particle Physics during his stay. The work of S.\,K.\ is supported by the DFG-funded Doctoral School KSETA.

\section*{Appendix}

In this appendix we provide some analytic expressions and numerical input parameters relevant for our analysis.

The loop function $L(x_x,x_j)$ is given by
\begin{equation}\label{eq:L}
L(x_{i},x_{j})=\left(\frac{x_{i}^{2}\log(x_{i})}{(x_{i}-x_{j})(1-x_{i})^{2}}+\frac{x_{j}^{2}\log(x_{j})}{(x_{j}-x_{i})(1-x_{j})^{2}}+\frac{1}{(1-x_{i})(1-x_{j})}\right)\,.
\end{equation}

\begin{table}[h!]
\centering
       \begin{tabular}{|l|l|}
       \hline 
$f_k = 155.9\,\text{MeV}$ &  $\eta_1=1.87$  \\
$\hat B_K=0.7625$ & $\eta_2=0.58$  \\
$f_D=209.2\,\text{MeV}$ & $\eta_3=0.49$ \\
{$ B_D(3\,\text{GeV})= 0.75$} & {$\eta_D(3\,\text{GeV}) = 0.77$}\\
$f_{B_d}\sqrt{\hat B_{B_d}} = 228\,\text{MeV}$ & $\eta_B =0.55$ \\
$f_{B_s}\sqrt{\hat B_{B_s}} = 275\,\text{MeV}$ & \\\hline
       \end{tabular}
\caption{Central values of QCD input parameters used in the numerical
analysis \cite{Aoki:2013ldr,Carrasco:2014uya,Buras:2001ra,Bazavov:2016nty}. }
\label{FlavourConstraintsTable}
\end{table}

The DM scattering amplitudes off protons, $f_p$,  and neutrons, $f_n$, receive various contributions discussed in \autoref{sec:direct}. Denoting the contributions from the DM-up (down) quark coupling by a superscript $u$ ($d$), we have:
\begin{itemize}
 \item tree level contributions
  \begin{equation}
  f^{\text{tree},u}_p=2f^{\text{tree},u}_n=\frac{|\lambda_{ut}|^2}{4m^2_\phi}\,, \qquad
2 f^{\text{tree},d}_p=f^{\text{tree},d}_n=\frac{|\tilde{\lambda}_{db}|^2}{4m^2_\phi}\,;
 \end{equation}
 \item box diagram contributions
 \begin{gather}
  f^{\text{box},u}_p=2f^{\text{box},u}_n=\sum_{i,j}\frac{|\lambda_{ui}|^2|\lambda_{jt}|^2}{32\pi^2m^2_\phi}L\left(\frac{m^2_{q_i}}{m^2_\phi},\frac{m^2_{\chi_j}}{m^2_\phi}\right)\,,\\
  2f^{\text{box},d}_p=f^{\text{box},d}_n=\sum_{i,j}\frac{|\tilde{\lambda}_{di}|^2|\tilde{\lambda}_{jb}|^2}{32\pi^2m^2_\phi}L\left(\frac{m^2_{q_i}}{m^2_\phi},\frac{m^2_{\chi_j}}{m^2_\phi}\right)\,,
 \end{gather}
with the loop function $L$ given in equation \eqref{eq:L};
 \item photon penguin contributions
 \begin{gather}
 f_p^{\text{photon},u}=-\sum_i\frac{|\lambda_{it}|^2e^2}{48\pi^2m^2_\phi}\left[\frac{3}{2}+\log\left(\frac{m^2_{q_i}}{m^2_\phi}\right)\right]\,,\\
 f_p^{\text{photon},d}=+\sum_i\frac{|\tilde{\lambda}_{ib}|^2e^2}{96\pi^2m^2_\phi}\left[\frac{3}{2}+\log\left(\frac{m^2_{q_i}}{m^2_\phi}\right)\right]\,;
 \end{gather}
 \item $Z$ penguin contributions:
\begin{eqnarray}
 f_p^{Z,u}&=& - \frac{3|\lambda_{tt}|^2e^2\left(\frac{1}{2}-2 \sin^2\theta_W \right)}{32\pi^2 \sin^2\theta_W \cos^2\theta_W m_Z^2}\frac{m^2_{t}}{m^2_\phi}\left[1+\log\left(\frac{m^2_{t}}{m^2_\phi}\right)\right]\,,\\
 f_n^{Z,u}&=& - \frac{3|\lambda_{tt}|^2e^2\left(-\frac{1}{2}\right)}{32\pi^2 \sin^2\theta_W \cos^2\theta_W m_Z^2}\frac{m^2_{t}}{m^2_\phi}\left[1+\log\left(\frac{m^2_{t}}{m^2_\phi}\right)\right]\,,
\end{eqnarray}
with $\theta_W$ being the weak mixing angle. {Note that due to the suppression by the squared quark mass, only the top contribution is relevant.}
\end{itemize}
$f_n$ and $f_p$ are then given by the sum of all these relevant contributions.

\bibliographystyle{JHEP}
\bibliography{sources}

\providecommand{\href}[2]{#2}\begingroup\raggedright\begin{thebibliography}{10}

\bibitem{Arcadi:2017kky}
G.~Arcadi, M.~Dutra, P.~Ghosh, M.~Lindner, Y.~Mambrini, M.~Pierre, S.~Profumo,
  and F.~S. Queiroz, {\it {The Waning of the WIMP? A Review of Models,
  Searches, and Constraints}},  \href{http://arxiv.org/abs/1703.07364}{{\tt
  arXiv:1703.07364}}.

\bibitem{Batell:2011tc}
B.~Batell, J.~Pradler, and M.~Spannowsky, {\it {Dark Matter from Minimal Flavor
  Violation}},  {\em JHEP} {\bf 08} (2011) 038,
  [\href{http://arxiv.org/abs/1105.1781}{{\tt arXiv:1105.1781}}].

\bibitem{DMFVPrimer}
P.~Agrawal, M.~Blanke, and K.~Gemmler, {\it {Flavored dark matter beyond
  Minimal Flavor Violation}},  {\em JHEP} {\bf 1410} (2014) 72,
  [\href{http://arxiv.org/abs/1405.6709}{{\tt arXiv:1405.6709}}].

\bibitem{Kilic:2015vka}
C.~Kilic, M.~D. Klimek, and J.-H. Yu, {\it {Signatures of Top Flavored Dark
  Matter}},  {\em Phys.Rev.} {\bf D91} (2015), no.~5 054036,
  [\href{http://arxiv.org/abs/1501.02202}{{\tt arXiv:1501.02202}}].

\bibitem{Agrawal:2011ze}
P.~Agrawal, S.~Blanchet, Z.~Chacko, and C.~Kilic, {\it {Flavored Dark Matter,
  and Its Implications for Direct Detection and Colliders}},  {\em Phys. Rev.}
  {\bf D86} (2012) 055002, [\href{http://arxiv.org/abs/1109.3516}{{\tt
  arXiv:1109.3516}}].

\bibitem{Cheung:2011zza}
K.~Cheung, K.~Mawatari, E.~Senaha, P.~Y. Tseng, and T.~C. Yuan, {\it {Top
  window for dark matter}},  {\em Int. J. Mod. Phys.} {\bf D20} (2011)
  1413--1421.

\bibitem{Kile:2011mn}
J.~Kile and A.~Soni, {\it {Flavored Dark Matter in Direct Detection Experiments
  and at LHC}},  {\em Phys. Rev.} {\bf D84} (2011) 035016,
  [\href{http://arxiv.org/abs/1104.5239}{{\tt arXiv:1104.5239}}].

\bibitem{Kamenik:2011nb}
J.~F. Kamenik and J.~Zupan, {\it {Discovering Dark Matter Through Flavor
  Violation at the LHC}},  {\em Phys. Rev.} {\bf D84} (2011) 111502,
  [\href{http://arxiv.org/abs/1107.0623}{{\tt arXiv:1107.0623}}].

\bibitem{Kumar:2013hfa}
A.~Kumar and S.~Tulin, {\it {Top-flavored dark matter and the forward-backward
  asymmetry}},  {\em Phys. Rev.} {\bf D87} (2013), no.~9 095006,
  [\href{http://arxiv.org/abs/1303.0332}{{\tt arXiv:1303.0332}}].

\bibitem{Chang:2013oia}
S.~Chang, R.~Edezhath, J.~Hutchinson, and M.~Luty, {\it {Effective WIMPs}},
  {\em Phys. Rev.} {\bf D89} (2014), no.~1 015011,
  [\href{http://arxiv.org/abs/1307.8120}{{\tt arXiv:1307.8120}}].

\bibitem{Kile:2013ola}
J.~Kile, {\it {Flavored Dark Matter: A Review}},  {\em Mod. Phys. Lett.} {\bf
  A28} (2013) 1330031, [\href{http://arxiv.org/abs/1308.0584}{{\tt
  arXiv:1308.0584}}].

\bibitem{Bai:2013iqa}
Y.~Bai and J.~Berger, {\it {Fermion Portal Dark Matter}},  {\em JHEP} {\bf 11}
  (2013) 171, [\href{http://arxiv.org/abs/1308.0612}{{\tt arXiv:1308.0612}}].

\bibitem{Batell:2013zwa}
B.~Batell, T.~Lin, and L.-T. Wang, {\it {Flavored Dark Matter and R-Parity
  Violation}},  {\em JHEP} {\bf 01} (2014) 075,
  [\href{http://arxiv.org/abs/1309.4462}{{\tt arXiv:1309.4462}}].

\bibitem{Agrawal:2014ufa}
P.~Agrawal, Z.~Chacko, and C.~B. Verhaaren, {\it {Leptophilic Dark Matter and
  the Anomalous Magnetic Moment of the Muon}},  {\em JHEP} {\bf 08} (2014) 147,
  [\href{http://arxiv.org/abs/1402.7369}{{\tt arXiv:1402.7369}}].

\bibitem{Agrawal:2014una}
P.~Agrawal, B.~Batell, D.~Hooper, and T.~Lin, {\it {Flavored Dark Matter and
  the Galactic Center Gamma-Ray Excess}},  {\em Phys. Rev.} {\bf D90} (2014),
  no.~6 063512, [\href{http://arxiv.org/abs/1404.1373}{{\tt arXiv:1404.1373}}].

\bibitem{Gomez:2014lva}
M.~A. Gomez, C.~B. Jackson, and G.~Shaughnessy, {\it {Dark Matter on Top}},
  {\em JCAP} {\bf 1412} (2014), no.~12 025,
  [\href{http://arxiv.org/abs/1404.1918}{{\tt arXiv:1404.1918}}].

\bibitem{Hamze:2014wca}
A.~Hamze, C.~Kilic, J.~Koeller, C.~Trendafilova, and J.-H. Yu, {\it
  {Lepton-Flavored Asymmetric Dark Matter and Interference in Direct
  Detection}},  {\em Phys. Rev.} {\bf D91} (2015), no.~3 035009,
  [\href{http://arxiv.org/abs/1410.3030}{{\tt arXiv:1410.3030}}].

\bibitem{Lee:2014rba}
C.-J. Lee and J.~Tandean, {\it {Lepton-Flavored Scalar Dark Matter with Minimal
  Flavor Violation}},  {\em JHEP} {\bf 04} (2015) 174,
  [\href{http://arxiv.org/abs/1410.6803}{{\tt arXiv:1410.6803}}].

\bibitem{Kile:2014jea}
J.~Kile, A.~Kobach, and A.~Soni, {\it {Lepton-Flavored Dark Matter}},  {\em
  Phys. Lett.} {\bf B744} (2015) 330--338,
  [\href{http://arxiv.org/abs/1411.1407}{{\tt arXiv:1411.1407}}].

\bibitem{Agrawal:2015kje}
P.~Agrawal, Z.~Chacko, E.~C. F.~S. Fortes, and C.~Kilic, {\it {Skew-Flavored
  Dark Matter}},  {\em Phys. Rev.} {\bf D93} (2016), no.~10 103510,
  [\href{http://arxiv.org/abs/1511.06293}{{\tt arXiv:1511.06293}}].

\bibitem{Lopez-Honorez:2013wla}
L.~Lopez-Honorez and L.~Merlo, {\it {Dark matter within the minimal flavour
  violation ansatz}},  {\em Phys. Lett.} {\bf B722} (2013) 135--143,
  [\href{http://arxiv.org/abs/1303.1087}{{\tt arXiv:1303.1087}}].

\bibitem{Blanke:2017tnb}
M.~Blanke and S.~Kast, {\it {Top-Flavoured Dark Matter in Dark Minimal Flavour
  Violation}},  {\em JHEP} {\bf 05} (2017) 162,
  [\href{http://arxiv.org/abs/1702.08457}{{\tt arXiv:1702.08457}}].

\bibitem{Jubb:2017rhm}
T.~Jubb, M.~Kirk, and A.~Lenz, {\it {Charming Dark Matter}},
  \href{http://arxiv.org/abs/1709.01930}{{\tt arXiv:1709.01930}}.

\bibitem{Chen:2015jkt}
M.-C. Chen, J.~Huang, and V.~Takhistov, {\it {Beyond Minimal Lepton Flavored
  Dark Matter}},  {\em JHEP} {\bf 02} (2016) 060,
  [\href{http://arxiv.org/abs/1510.04694}{{\tt arXiv:1510.04694}}].

\bibitem{Buras:2000dm}
A.~J. Buras, P.~Gambino, M.~Gorbahn, S.~Jager, and L.~Silvestrini, {\it
  {Universal unitarity triangle and physics beyond the standard model}},  {\em
  Phys. Lett.} {\bf B500} (2001) 161--167,
  [\href{http://arxiv.org/abs/hep-ph/0007085}{{\tt hep-ph/0007085}}].

\bibitem{D'Ambrosio:2002ex}
G.~D'Ambrosio, G.~F. Giudice, G.~Isidori, and A.~Strumia, {\it {Minimal flavor
  violation: An Effective field theory approach}},  {\em Nucl. Phys.} {\bf
  B645} (2002) 155--187, [\href{http://arxiv.org/abs/hep-ph/0207036}{{\tt
  hep-ph/0207036}}].

\bibitem{Buras:2003jf}
A.~J. Buras, {\it {Minimal flavor violation}},  {\em Acta Phys. Polon.} {\bf
  B34} (2003) 5615--5668, [\href{http://arxiv.org/abs/hep-ph/0310208}{{\tt
  hep-ph/0310208}}].

\bibitem{Blanke:2006xr}
M.~Blanke, A.~J. Buras, A.~Poschenrieder, S.~Recksiegel, C.~Tarantino,
  S.~Uhlig, and A.~Weiler, {\it {Another look at the flavour structure of the
  littlest Higgs model with T-parity}},  {\em Phys. Lett.} {\bf B646} (2007)
  253--257, [\href{http://arxiv.org/abs/hep-ph/0609284}{{\tt hep-ph/0609284}}].

\bibitem{Bona:2016bvr}
{\bf UTfit} Collaboration, M.~Bona, {\it {Unitarity Triangle analysis beyond
  the Standard Model from UTfit}},  {\em PoS} {\bf ICHEP2016} (2016) 149.

\bibitem{Papucci:2014iwa}
M.~Papucci, A.~Vichi, and K.~M. Zurek, {\it {Monojet versus the rest of the
  world I: t-channel models}},  {\em JHEP} {\bf 11} (2014) 024,
  [\href{http://arxiv.org/abs/1402.2285}{{\tt arXiv:1402.2285}}].

\bibitem{Aad:2014wea}
{\bf ATLAS} Collaboration, G.~Aad et~al., {\it {Search for squarks and gluinos
  with the ATLAS detector in final states with jets and missing transverse
  momentum using $\sqrt{s}=8$ TeV proton--proton collision data}},  {\em JHEP}
  {\bf 09} (2014) 176, [\href{http://arxiv.org/abs/1405.7875}{{\tt
  arXiv:1405.7875}}].

\bibitem{Hurth:2009ke}
T.~Hurth and W.~Porod, {\it {Flavour violating squark and gluino decays}},
  {\em JHEP} {\bf 08} (2009) 087, [\href{http://arxiv.org/abs/0904.4574}{{\tt
  arXiv:0904.4574}}].

\bibitem{Blanke:2013zxo}
M.~Blanke, G.~F. Giudice, P.~Paradisi, G.~Perez, and J.~Zupan, {\it {Flavoured
  Naturalness}},  {\em JHEP} {\bf 06} (2013) 022,
  [\href{http://arxiv.org/abs/1302.7232}{{\tt arXiv:1302.7232}}].

\bibitem{Agrawal:2013kha}
P.~Agrawal and C.~Frugiuele, {\it {Mixing stops at the LHC}},  {\em JHEP} {\bf
  01} (2014) 115, [\href{http://arxiv.org/abs/1304.3068}{{\tt
  arXiv:1304.3068}}].

\bibitem{Arana-Catania:2014ooa}
M.~Arana-Catania, S.~Heinemeyer, and M.~J. Herrero, {\it {Updated Constraints
  on General Squark Flavor Mixing}},  {\em Phys. Rev.} {\bf D90} (2014), no.~7
  075003, [\href{http://arxiv.org/abs/1405.6960}{{\tt arXiv:1405.6960}}].

\bibitem{Backovic:2015rwa}
M.~Backovi\'c, A.~Mariotti, and M.~Spannowsky, {\it {Signs of Tops from Highly
  Mixed Stops}},  {\em JHEP} {\bf 06} (2015) 122,
  [\href{http://arxiv.org/abs/1504.00927}{{\tt arXiv:1504.00927}}].

\bibitem{Blanke:2015ulx}
M.~Blanke, B.~Fuks, I.~Galon, and G.~Perez, {\it {Gluino Meets Flavored
  Naturalness}},  {\em JHEP} {\bf 04} (2016) 044,
  [\href{http://arxiv.org/abs/1512.03813}{{\tt arXiv:1512.03813}}].

\bibitem{Bona:2005eu}
{\bf UTfit} Collaboration, M.~Bona et~al., {\it {The UTfit collaboration report
  on the status of the unitarity triangle beyond the standard model. I.
  Model-independent analysis and minimal flavor violation}},  {\em JHEP} {\bf
  03} (2006) 080, [\href{http://arxiv.org/abs/hep-ph/0509219}{{\tt
  hep-ph/0509219}}].

\bibitem{Bona:2007vi}
{\bf UTfit} Collaboration, M.~Bona et~al., {\it {Model-independent constraints
  on $\Delta F=2$ operators and the scale of new physics}},  {\em JHEP} {\bf
  03} (2008) 049, [\href{http://arxiv.org/abs/0707.0636}{{\tt
  arXiv:0707.0636}}].

\bibitem{Amhis:2016xyh}
Y.~Amhis et~al., {\it {Averages of $b$-hadron, $c$-hadron, and $\tau$-lepton
  properties as of summer 2016}},  \href{http://arxiv.org/abs/1612.07233}{{\tt
  arXiv:1612.07233}}.

\bibitem{Steigman:2012nb}
G.~Steigman, B.~Dasgupta, and J.~F. Beacom, {\it {Precise Relic WIMP Abundance
  and its Impact on Searches for Dark Matter Annihilation}},  {\em Phys.Rev.}
  {\bf D86} (2012) 023506, [\href{http://arxiv.org/abs/1204.3622}{{\tt
  arXiv:1204.3622}}].

\bibitem{Ade:2015xua}
{\bf Planck} Collaboration, P.~A.~R. Ade et~al., {\it {Planck 2015 results.
  XIII. Cosmological parameters}},  \href{http://arxiv.org/abs/1502.01589}{{\tt
  arXiv:1502.01589}}.

\bibitem{Aprile:2017iyp}
{\bf XENON} Collaboration, E.~Aprile et~al., {\it {First Dark Matter Search
  Results from the XENON1T Experiment}},
  \href{http://arxiv.org/abs/1705.06655}{{\tt arXiv:1705.06655}}.

\bibitem{Akerib:2016vxi}
D.~S. Akerib et~al., {\it {Results from a search for dark matter in the
  complete LUX exposure}},  \href{http://arxiv.org/abs/1608.07648}{{\tt
  arXiv:1608.07648}}.

\bibitem{Diglio:2016stt}
{\bf XENON} Collaboration, S.~Diglio, {\it {XENON1T: the start of a new era in
  the search for Dark Matter}},  {\em PoS} {\bf DSU2015} (2016) 032.

\bibitem{Akerib:2015cja}
{\bf LZ} Collaboration, D.~S. Akerib et~al., {\it {LUX-ZEPLIN (LZ) Conceptual
  Design Report}},  \href{http://arxiv.org/abs/1509.02910}{{\tt
  arXiv:1509.02910}}.

\bibitem{Aalbers:2016jon}
{\bf DARWIN} Collaboration, J.~Aalbers et~al., {\it {DARWIN: towards the
  ultimate dark matter detector}},  {\em JCAP} {\bf 1611} (2016), no.~11 017,
  [\href{http://arxiv.org/abs/1606.07001}{{\tt arXiv:1606.07001}}].

\bibitem{Liu:2016cnk}
H.~Liu, T.~R. Slatyer, and J.~Zavala, {\it {Contributions to cosmic
  reionization from dark matter annihilation and decay}},  {\em Phys. Rev.}
  {\bf D94} (2016), no.~6 063507, [\href{http://arxiv.org/abs/1604.02457}{{\tt
  arXiv:1604.02457}}].

\bibitem{Aoki:2013ldr}
S.~Aoki, Y.~Aoki, C.~Bernard, T.~Blum, G.~Colangelo, et~al., {\it {Review of
  lattice results concerning low-energy particle physics}},  {\em Eur.Phys.J.}
  {\bf C74} (2014) 2890, [\href{http://arxiv.org/abs/1310.8555}{{\tt
  arXiv:1310.8555}}].

\bibitem{Carrasco:2014uya}
N.~Carrasco, M.~Ciuchini, P.~Dimopoulos, R.~Frezzotti, V.~Gimenez, et~al., {\it
  {$D^0$$-\bar{D}^0$ mixing in the standard model and beyond from $N_f$ =2
  twisted mass QCD}},  {\em Phys.Rev.} {\bf D90} (2014), no.~1 014502,
  [\href{http://arxiv.org/abs/1403.7302}{{\tt arXiv:1403.7302}}].

\bibitem{Buras:2001ra}
A.~J. Buras, S.~J{\"a}ger, and J.~Urban, {\it {Master formulae for $\Delta F=2$
  NLO QCD factors in the standard model and beyond}},  {\em Nucl. Phys.} {\bf
  B605} (2001) 600--624, [\href{http://arxiv.org/abs/hep-ph/0102316}{{\tt
  hep-ph/0102316}}].

\bibitem{Bazavov:2016nty}
{\bf Fermilab Lattice, MILC} Collaboration, A.~Bazavov et~al., {\it
  {$B^0_{(s)}$-mixing matrix elements from lattice QCD for the Standard Model
  and beyond}},  {\em Phys. Rev.} {\bf D93} (2016), no.~11 113016,
  [\href{http://arxiv.org/abs/1602.03560}{{\tt arXiv:1602.03560}}].

\end{thebibliography}\endgroup

\end{document}